\documentclass[preprint,authoryear,letterpaper]{elsarticle}
\usepackage{natbib}
\bibstyle{aa}
\bibpunct{(}{)}{,}{a}{}{;}
\usepackage{epsfig}
\usepackage{amssymb}
\usepackage{morefloats}
\usepackage[font=normal]{caption}

\newcommand{\kg}{\mbox{$\rm\,kg$}}

\newcommand{\myr}{\mbox{$\rm\,Myr$}}

\newcommand{\yr}{\mbox{$\rm\,yr$}}

\newcommand{\hour}{\mbox{$\rm\,h$}}
\newcommand{\arcdeg}{^{\circ}}

\newcommand{\km}{\mbox{$\rm\,km$}}
\newcommand{\m}{\mbox{$\rm\,m$}}

\newcommand{\au}{\mbox{$\rm\,AU$}}

\newcommand{\tiunits}{\mbox{$\rm\,J\,m^{-1}\,s^{-1/2}\,K^{-1}$}}
\newcommand{\beq}{\begin{equation}}
\newcommand{\eeq}{\end{equation}}
\newcommand{\po}{\phantom{1}}

\begin{document}

\begin{frontmatter}


\title{Obliquities of ``Top-Shaped'' Asteroids May Not Imply Reshaping by
YORP Spin-up}
\author[ou,umd]{Thomas S.~Statler\corref{cor}}
\ead{statler@ohio.edu}
\address[ou]{Astrophysical Institute, Department of Physics and Astronomy, 251B
Clippinger Research Laboratories, Ohio University, Athens, OH 45701, USA}
\address[umd]{Department of Astronomy, University of Maryland, College Park, MD,
20742, USA}


\begin{abstract}
The timescales over which the YORP effect alters the rotation period and the
obliquity of a small asteroid can be very different, because the corresponding
torques couple to different aspects of the object's shape. For nearly
axisymmetric, ``top-shaped'' near-Earth asteroids such as 101955 Bennu,
spin timescales are an order of magnitude or more longer than obliquity
timescales, which are $\sim 10^5$ to $10^6 \yr$. The observed low
obliquities (near $0\arcdeg$ or $180\arcdeg$) of top-shaped asteroids
do not constitute evidence that they acquired their present shapes and spins
through YORP spin-up, because low obliquities are expected regardless
of the spin-up or reshaping mechanism.
\end{abstract}
 

\end{frontmatter}


\section{Introduction}

High-resolution radar observations of near-Earth asteroids (NEAs) have
produced a wealth of information about their surface properties, rotation
rates, and 3-dimensional shapes \citep{OstroReview}. One intriguing result
from these studies is the recurring emergence of shapes reminiscent of a
child's top---nearly axisymmetric, slightly to moderately oblate, with an
elevated ridge around the equator. The dozen or so top-shaped objects
identified to date tend to have rapid rotation and axial obliquities
near $0\arcdeg$ or $180\arcdeg$ (collectively referred to here as
``low obliquity''), and several have satellites
\citep{Ost06kw4,Bus07da,Bus11ev5,Nol13bennu,Bro11cc}.

One possible explanation for the formation of top-shapes, both with and
without companions, was offered by \citet{WalshNature,Wal12}.
These authors simulated the dynamical evolution of idealized rubble piles
composed of spheres that interact through dissipative two-body collisions.
They subjected the rubble piles to a steady increase in angular momentum,
ostensibly arising from radiation recoil torques
\citep[the YORP effect;][]{Pad69,Rub00}. They found that
some of the objects evolved, through centrifugally driven movement of
surface material, to top shapes. By continuing to add angular
momentum, they could force the tops to shed mass, which reaccumulated in orbit
to make satellites. \citet{WalshNature} highlighted the strong similarity of
their best results to the well-studied object 1999~KW$_4$, establishing
YORP spin-up as a likely candidate mechanism.

These simulations are so visually compelling that 
YORP is now commonly invoked as {\em the only\/} mechanism
responsible for top-shapes. \citet{Kel10} state that the shape of the
main-belt asteroid 2687 Steins is ``probably the result of reshaping
due to [YORP] spin-up''; \citet{Bus11ev5} describe 1999~KW$_4$'s equatorial
ridge and satellite as ``believed to have formed due to YORP spin-up\dots'';
and \citet{Wal12b} cite YORP-induced ``bulk reshaping'' as ``the
cause for the ubiquitous `top-shape' and equatorial ridge.''

Given the current state of knowledge, however, uncritical acceptance of
the YORP spin-up mechanism as the only option for the formation of
top shapes is logically unwise, for the following reasons:
\begin{enumerate}
\item
Not all spin-accelerated rubble piles become tops, and it is not yet
determined whether the properties of those that do correspond to the
properties of real objects. \citet{Wal12} found that objects with low
initial angles of friction $\phi$ evolved, not to tops, but to highly
triaxial or prolate bodies \citep[see also][]{Hol10,Jac11,TanACM14,Cot14}.
Terrestrial materials like gravel or sand have larger values
of $\phi$ ($\gtrsim 20 \arcdeg$), but it is by no means established
that asteroidal materials will behave similarly. The high-$\phi$ objects of
\citet{WalshNature}, which did become tops,
were initialized in hexagonal-close-pack configuration,
which provides extra rigidity by making motion of material below the surface
impossible unless the object expands and the bulk density decreases.
Furthermore, light curve observations show an abundance of rapidly
rotating objects in the few-km size range that are significantly
non-axisymmetric \citep{Pra00,LCDB}.
\item
YORP is not a limitless source of angular momentum. YORP
spin-up will weaken as an object being reshaped becomes more
symmetric. For objects having reflection symmetry (including
axisymmetric oblate or prolate spheroids as well as triaxial ellipsoids) and
rotating about a principal axis, the secular YORP effect on spin is identically
zero.\footnote{The arguments in sections 1 and 2 of this paper make use of
the standard assumption that the recoil force from thermal photons is normal
to the radiating surface. The consequences of loosening this assumption
are discussed in
section 3.}  Small deviations from symmetry are equally likely to produce
positive or negative spin torques; hence gradual reshaping may lead to
a stochastic random walk in spin rate \citep{StatlerYORP} and/or to
YORP self-limitation \citep{Cot14}, either of which could arrest reshaping
and prevent mass shedding or fission. This scenario differs qualitatively
from the continual spin-up assumed in the simulations.
\item
Other mechanisms that may also reshape and/or accelerate the spins
of rubble piles have not been ruled out. Leading contenders are
disruptive impacts \citep{Lei00,Kor06} and
catastrophic disruptions followed by reaccumulation \citep{Mic13}.
Tidal torques in close planetary encounters may also contribute but
are expected to play a lesser role \citep{Wal06,Wal08}.
\end{enumerate}

A tempting argument to invoke in support of the YORP spin-up model
uses the tendency for tops and binaries to have obliquities $\epsilon$ close
to $0\arcdeg$ or $180\arcdeg$, which have been identified with stable end
states of the YORP cycle \citep{Cap04}. The argument is that, since tops are
found near these end states, YORP must have been in operation for longer
than the characteristic YORP timescale, over which time YORP must have
significantly modified the spins. \citet{PraACM14} applies this argument
to the general population of binary asteroids; \citet{Pol14} employs a form
of it in his discussion of asteroid pairs.

The point of this {\em Note\/} is to show that, at least for the top-shaped
asteroids, the argument is fallacious. This is because the timescale for
YORP to change the orientation of an object may have nothing to do with
the timescale over which it changes the spin rate; and for nearly symmetric
objects the latter timescale can be an order of magnitude or more longer than
the former. The low obliquities of tops do not imply that they acquired their
present shapes and spins through YORP spin-up, because low obliquities are
expected regardless of the spin-up or reshaping mechanism.

\section{YORP Evolution of Symmetric and Nearly Symmetric Asteroids}

The essential property of YORP in this discussion is that the torque
component that changes the spin rate and the components that change the
axis orientation couple, at leading order, to different attributes of the
surface. The spin torque couples to chirality---the difference between
eastward and westward facing slopes---while the other components couple merely
to asphericity. (Mathematically, this concerns the symmetric and
antisymmetric terms in the Fourier expansion of the topography:
the spin torque couples only to the antisymmetric terms, the orientation
component to the symmetric terms.) Thus, even axisymmetric spheroids,
which have zero spin torque, will have their axes reoriented by YORP,
and will have their obliquities changed if they have finite
thermal inertia $\Gamma$. These results 
have been derived analytically by \citet{Bre07,Bre08}, and
\citet{Kaa13}, but seem to have been underappreciated, perhaps owing
to the highly mathematical presentations in those papers.

\begin{figure}[t]
\begin{center}
\includegraphics[width=3.9in]{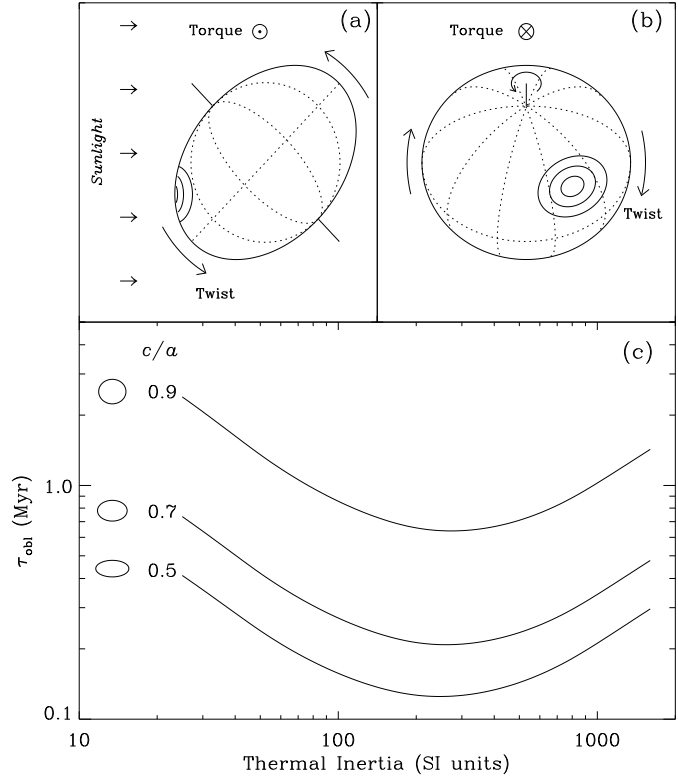}
\caption{YORP torque components on an axisymmetric oblate spheroid.
($a$) An object with zero thermal inertia is shown at its northern summer
solstice. Re-radiation from the sub-solar warm spot ({\it contours\/})
produces a twist in the direction indicated, i.e., a torque directed
out of the page that drives a precession of the rotation axis.
($b$) Sun's-eye view of the same body, with finite thermal inertia,
rotating counterclockwise. The warm spot is displaced eastward along a
parallel of latitude. The recoil force has a downward component in the plane of
the page, resulting in a clockwise twist and a torque into the page that
lowers (raises) the obliquity of a direct (retrograde) rotator toward
$\epsilon = 0\arcdeg$ ($180\arcdeg$).
($c$) Characteristic YORP timescale, $\tau_{\rm obl}$, for obliquity to
evolve from $60\arcdeg$ to $5\arcdeg$, {\em vs}. thermal inertia $\Gamma$
for fiducial, black oblate spheroids of density $1000 \kg \m^{-3}$ and
mass $5.346 \times 10^{11} \kg$, with $2 \hour$ rotation periods and on
circular orbits of radius $1 \au$. Curves correspond to different values
of short-to-long axis ratio (polar flattening) as indicated.
Any rotating oblate spheroid with $\Gamma>0$ will evolve toward
$\epsilon = 0\arcdeg$ or $180\arcdeg$, regardless of what made it oblate.
\label{f.torques}
}
\end{center}
\end{figure}

Figure \ref{f.torques} illustrates the origin of the axis-reorienting torque
components on an axisymmetric oblate spheroid, for which the spin torque
is identically zero at all times. Figure \ref{f.torques}a shows the case of a
body with $\Gamma=0$, at its northern summer solstice, illuminated
by sunlight from the left. A warm spot is generated, centered around
the sub-solar point, from which thermal re-radiation produces a recoil force
normal to the surface. As the force is not directed toward the center of
mass, the result is a twist in the direction indicated, corresponding to a
torque directed out of the page. Half an orbit later, with the illumination
from the right, the recoil twist is in the same sense, adding to the secular
effect. For $\Gamma=0$, this is the only non-zero component of torque, and it
drives a precession of the rotation axis about the orbit normal.
Figure \ref{f.torques}b shows a Sun's-eye view of the same body, now with
$\Gamma>0$. As a result of heat conduction and rotation
(counterclockwise looking down on the north pole, as indicated),
the warm spot is carried eastward, approximately along a parallel of
latitude. The recoil force now has a component pointing downward in the
diagram. The downward push on the right side of the body results in a
clockwise twist and a torque directed into the page. Half an orbit later,
illumination from behind creates an upward push on the left side of the
body and a torque in the same direction. This contribution to the secular
torque acts to lower the obliquity of a direct rotator toward $\epsilon =
0\arcdeg$, and raise the obliquity of a retrograde rotator toward $\epsilon =
180\arcdeg$.

As a measure of the characteristic timescale for this axis-righting
process, I define $\tau_{\rm obl}$ as the time for the obliquity of a
direct rotator
to evolve from $60\arcdeg$ (the median value for rotation poles
distributed randomly over one hemisphere) to $5\arcdeg$
(a typical observational uncertainty for well-determined rotation poles). Figure
\ref{f.torques}c shows this timescale for three fiducial oblate spheroids over
a range of thermal inertias, computed using the thermophysical
code TACO \citep{StatlerYORP}. These fiducial objects are black (zero Bond
albedo $A$ and unit blackbody radiative efficiency $\epsilon_{\rm bb}$),
with a uniform density of
$\rho = 1000 \kg \m^{-3}$ and volumes equal to that of a sphere $D=1\km$ in
diameter, differing only in their polar-to-equatorial axis ratio. They are
assumed to be on circular ($e=0$) orbits of semi-major axis $a=1\au$,
rotating about their short axes with period $P=2\hour$.
Thermal inertias range from small values ($\sim 10 \tiunits)$ characteristic
of fine regolith, through intermediate values ($\sim 10^2 \tiunits$)
characteristic of fractured rock, to high values ($> 10^3 \tiunits$)
typical of monolithic rock.
The figure shows that righting times are short:
$\tau_{\rm obl} < 1\myr$ for all likely values of $\Gamma$, even for objects
that are only 10\% aspherical, and are in the realm of $0.1 \myr$ for moderate
flattenings and thermal inertias characteristic of fractured rock.
Axis-righting occurs at constant spin rate, since the spin component of
torque is zero and the timescale for YORP spin-up or spin-down is infinite.
Any rotating oblate spheroid will evolve toward $0\arcdeg$ or $180\arcdeg$
obliquity, regardless of what made it oblate.
(The results can be scaled to other objects and orbits using the relation
$\tau_{\rm obl} \propto \rho D^2 a^2 (1-e^2)^{1/2} P^{-1}$ and rescaling
$\Gamma$ so that $\Gamma P^{-1/2} = \mbox{\rm constant}$. For Lambertian
reflection and emission, to leading order
$\tau_{\rm obl} \propto \left[ \epsilon_{\rm bb} (1-A)\right]^{-1}$.)

Real objects are not precisely symmetric, and will have nonzero spin torques
owing to deviations from reflection symmetry. But for top-shaped
objects the spin torque is still typically an order of magnitude
smaller than the obliquity torque. To demonstrate, I calculate
the YORP effect on four well-observed objects with high-resolution radar
models: 101955 Bennu, (29075) 1950~DA, (341843) 2008~EV$_5$, and (66391)
1999~KW$_4$. The adopted parameters are given in Table \ref{t.objects}.
For 2008~EV$_5$, $\Gamma$ has been estimated from thermal
infrared observations \citep{Ali14}. Its density is poorly constrained;
\citet{Bus11ev5} give $3000 \kg \m^{-3}$ as an upper limit, which I
adopt as a conservative estimate but which is higher
than expected for a rubble pile.  For Bennu and 1950~DA, observations
of Yarkovsky drift \citep{Che14,Eme14,Roz14da} permit constraints on
$\rho$ and $\Gamma$. For 1999~KW$_4$, no information on $\Gamma$
is available, and so I simply adopt $100 \tiunits$ as an intermediate 
value. The torque calculation includes 1-dimensional
heat conduction with the full nonlinear radiative boundary condition at the
surface, as well as self-heating by reflected sunlight and thermal emission.
Only the shallow diurnal thermal
wave is calculated. The seasonal effect (which vanishes for zero obliquity)
is neglected, as are surface roughness and beaming effects. I also assume
for simplicity that the object remains in its present orbit for the length
of the calculation.

\begin{table}[t]
{\centering
\caption{Adopted Physical Parameters for Modeled Asteroids\label{t.objects}}
\begin{tabular}{llccccccc}
\hline
Name & Shape Model & $\rho$ & $\Gamma$ & $A$ & $\epsilon$ & $P$ & $a$ & $e$ \\
\hline
101955 Bennu & \citet{Nol13bennu} & $1260^{\rm a}$ & $310^{\rm b}$ & $0.02^{\rm a}$ & $175^{\rm a}$ & $15437$ & $1.126$ & $0.024$ \\
(29075) 1950 DA & \citet{Bus07da} & $1700^{\rm c}$ & $\po 24^{\rm c}$ & $0.20^{\rm c}$ & $168^{\rm c}$ & \po $7638$ & $1.699$ & $0.508$ \\
(341843) 2008 EV$_5$ & \citet{Bus11ev5} & $3000^{\rm d}$ & $450^{\rm e}$ & $0.12^{\rm d}$ & $175^{\rm d}$ & $13410$ & $0.958$ & $0.081$ \\
(66391) 1999 KW$_4$ & \citet{Ost06kw4} & $2081^{\rm f}$ & $100\po$ & $0.20^{\rm g}$ & $\po\po 3^{\rm f}$ & $\po 9952$ & $0.642$ & $0.689$ \\
\hline\\[-.5em]
\end{tabular}\par}
{\em Note:\/} Columns list object name, source for shape model,
bulk density $\rho$ in $\kg \m^{-3}$, thermal inertia $\Gamma$ in
$\tiunits$, Bond albedo $A$, obliquity $\epsilon$ in degrees, rotation
period $P$ in s, orbital semi-major axis $a$ in $\au$,
and orbital eccentricity $e$. Additional data sources:
$^{\rm a}$\citet{Che14}; $^{\rm b}$\citet{Eme14}; $^{\rm c}$\citet{Roz14da};
$^{\rm d}$\citet{Bus11ev5}; $^{\rm e}$\citet{Ali14}; $^{\rm f}$\citet{Fah08kw4};
$^{\rm g}${\tt
http://echo.jpl.nasa.gov/$\sim$lance/asteroid\underbar{~}radar\underbar{~}properties/nea.radaralbedo.html}.
$P$, $a$, and $e$ values are from the JPL Small Body Database
({\tt http://ssd.jpl.nasa.gov/sbdb.cgi}).
\end{table}

For the following discussion, I adopt a working premise that is intentionally
counter to the YORP spin-up concept: I assume that each of these objects
attained its current shape through some unspecified process or event some time
in the past, and has retained that shape until now. I integrate the coupled
spin and obliquity evolution driven by YORP backward in time, from the present
period and obliquity, and ask how long it should have taken for the object
to reach its current spin state. The initial obliquity, at the time of the
shape-setting event, is, of course, unknown.

\begin{figure}[t]
\begin{center}
\includegraphics[width=4.4in]{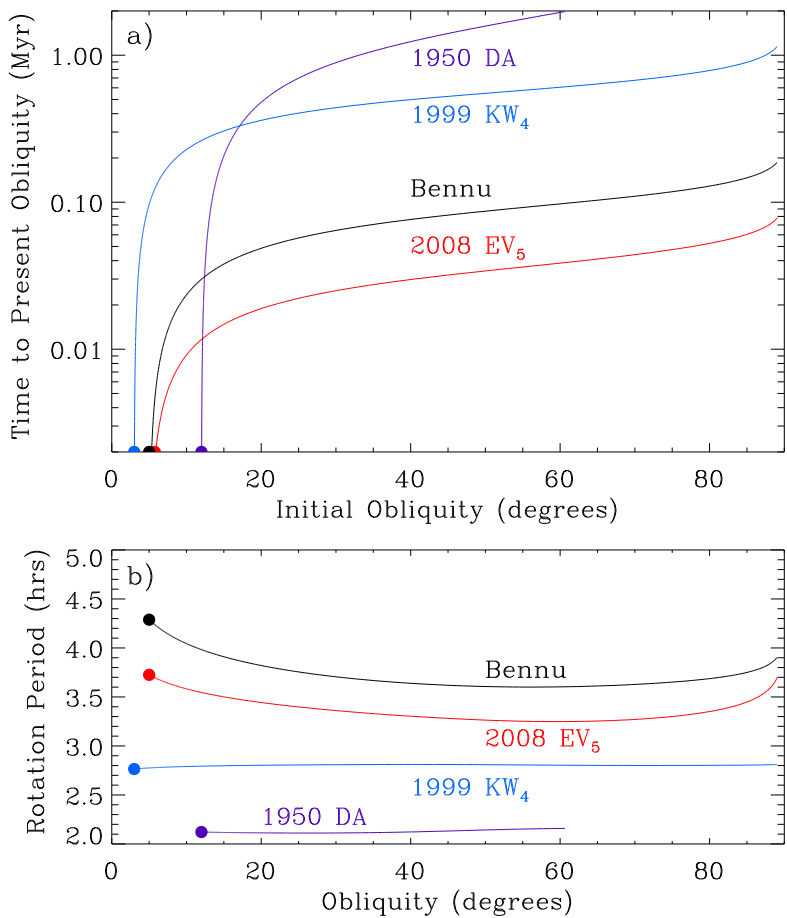}
\caption{($a$, top) Time required for 4 top-shaped asteroids to reach their
current spin states by YORP evolution, as a function of the initial obliquities
at the times their present shapes were set.
($b$, bottom) Evolution in obliquity and rotation period. Axis-righting
would occur in $\lesssim 1 \myr$, at nearly constant spin rate. (In both
panels, the complement of the obliquity, rather than the obliquity, is
plotted for the retrograde rotators.) 
\label{f.real_objects}
}
\end{center}
\end{figure}

Figure \ref{f.real_objects}a shows the times to reach the current spin states,
as a function of initial obliquity. The black curve shows that Bennu could
have reached its present obliquity in, at most, $0.2 \myr$; the median time,
assuming a statistical ensemble of random, isotropically oriented initial
rotation poles, would be $0.1 \myr$. For 1950~DA, the larger mass and
low $\Gamma$ make the timescale longer: to reach its current
state, 1950~DA would require a median time of $2 \myr$. 
1999~KW$_4$ and 2008~EV$_5$ would reach their present orientations in
median times of $0.6 \myr$ and $0.04 \myr$, respectively.
These times are all significantly shorter than the
$10\myr$ median lifetimes of NEAs \citep{Gla00}, and much shorter
than the $\sim 100\myr$ time between large impacts that significantly
change the magnitude or direction of the angular momentum
\citep{Far92,Far98,Mor03,Mar11}.

During this reorientation, the spin rates would have changed only modestly.
Figure \ref{f.real_objects}b shows evolutionary
tracks in obliquity and rotation period, terminating at their current values.
While it is impossible to definitively calculate the spin torque
without knowing the internal mass distribution and the
topography at much higher resolution than the radar models
allow, the tracks indicate that all four objects would have reoriented to
their present obliquities with little change in period. 2008~EV$_5$ and
Bennu would have slowed by at most 12\% and 20\%, respectively, if they
started from $\epsilon \approx 60\arcdeg$. 1999~KW$_4$ and 1950~DA, behaving
much more like symmetric spheroids, would have evolved at nearly constant
spin rate.

\section{Discussion}

Axis righting by YORP does not imply spin-up by YORP, since obliquity and
spin timescales can be very different for shapes that are close to spheroidal.
The above results highlight the point made in Section 1, that once an asteroid
acquires a near-symmetric shape, YORP is not a particularly effective supplier
of angular momentum. Furthermore, alternative mechanisms for spin-up or
reshaping that have no intrinsic orientation bias are not ruled out
by low present-day obliquities, because YORP would quickly re-orient objects
to low obliquity anyway. {\em Regardless of what gave these objects their
characteristic shapes or rapid spins, we would expect to find them today
with low obliquities, spinning at close to their original speeds.}

The leading alternative to YORP spin-up, for reshaping as well
as for making binaries, is disruptive impacts. The outcomes of
impacts depend sensitively on impact parameter \citep{Lei00}, rotation
\citep{Bal14}, and physical makeup of the target body
\citep{Kor06,Mic04,Jut10}, as well as on impact energy. Oblate
remnants with rotation periods $P < 8\hour$ and small satellites tend to
result from intermediate angular momentum impacts \citep{Lei00}, and
from rubble piles composed of same-sized pieces
\citep{Kor06}.  When the target is fully dissociated and the remains
reaccumulate \citep{Mic13}, preliminary simulations
suggest it can also lead to rapidly rotating, near-oblate remnants
(P.\ Michel, private communication). The binary YORP effect
\cite[BYORP;][]{Cuk05}, as well as tidal effects, can
cause satellites to migrate inward and collapse onto the primary
\citep{Tay11,Tay14}; accretion of orbiting material might then lead to the
buildup of an equatorial ridge \citep{Bus11ev5}.

Figure \ref{f.real_objects} does not constrain
the ages of top-shaped NEAs or the times since reshaping.
The point is not that these objects were reshaped or
acquired their spins only $\sim 0.1 \myr$ ago; merely that {\em whenever\/}
this occurred, once in the near-Earth region they would have been quickly
reoriented at nearly constant spin rates.
Impact-driven reshaping would be expected to occur
predominantly in the main belt and among the NEAs with $a>2\au$ \citep{Gla97}.
Such objects could have been reshaped
well before migrating to more Earthlike orbits where they would have been
quickly reoriented.

1999~KW$_4$ is a binary, as are some other top-shaped NEAs including
2013~WT$_{44}$ (P.\ Taylor, private communication) and 1994~CC
\citep[which is a trinary;][]{Bro11cc}. It is easily shown that YORP-driven
axis righting is an adiabatic change for the binary orbit, and therefore
the orbit pole will be ``dragged along'' and
reoriented in the same way. Using parameters from \citet{Fah08kw4}
for 1999~KW$_4$, I calculate the nodal precession
time for a test particle at the distance
of the secondary to be 110 days.
\citet{Fah08kw4} show that 1999~KW$_4$ is in a Cassini state, in which the
primary's spin axis and the secondary's orbit pole precess together with a
90-day period. This is close to the test-particle result, and some 6 orders
of magnitude shorter than the YORP axis-righting timescale, hence the latter
is adiabatic.

Photometrically identified NEA binaries predominantly have small
($D < 4 \km)$ primaries with
rotation periods $2\hour < P < 5 \hour$ and light curve amplitudes
$\lesssim0.2$ mag, implying shapes not very far from axisymmetry
\citep{Pra06}. Their polar to equatorial axis ratios $c/a$ are not well
constrained by light curves, but if they are moderately flattened
($c/a \lesssim 0.9$), then the above results should apply to them
as well. That is, regardless of the process or processes that made them 
oblate and gave them satellites, we would expect YORP to have efficiently
aligned their spin and binary orbit poles with their heliocentric orbit poles.

For inner main belt binaries, a preference for such alignments has been
observed by \citet{Pra12}. They propose that this is a result
of YORP reorientation, before, during, or after formation, and estimate a
reorientation rate for a $D=4\km$ object equivalent to
$\tau_{\rm obl} \approx 140\myr$. The results of Section 2 actually suggest
a shorter time; scaling the curve for $c/a=0.7$ in Figure \ref{f.torques}c
to the average parameters in Table 1 of \citet{Pra12} and adopting their
assumed $\rho=2500\kg\m^{-3}$, I find that $\tau_{\rm obl}$
can be as low as $30\myr$ if $\Gamma \approx 300\tiunits$, which
is a reasonable thermal inertia for a surface composed of broken rocks.

None of the arguments presented here excludes YORP from playing significant
roles in asteroid reshaping or binary formation. Collisional spin-up need
not dominate YORP spin-up in general. YORP could be responsible
for the majority of fast-rotating non-axisymmetric asteroids, and impacts
for the minority of objects that are tops. YORP spin-up of asymmetric
objects may also produce some binaries by fission, without passing through an
axisymmetric phase. Whether the progenitor object is first forced to low
obliquity will depend on the details of the evolution. Righting times for
prolate spheroids are approximately twice those for the corresponding oblate
spheroids (Fig.\ \ref{f.torques}c); but the sequence of shape changes
that would inevitably occur before fission may result
in stochastic YORP evolution in both obliquity and spin rate \citep{Cot14}.
YORP may also continue to modify the spins of objects that have been
shaped by other processes. Near-axisymmetric objects would be particularly
prone to reversals in spin torque caused by small mass movements.
The resulting YORP self-limitation \citep{Cot14} may prevent substantial
spin-up {\em or\/} spin-down, keeping the rotation rate high and
occasionally driving material off the surface.

Ultimately, YORP {\em may\/} prove to be the dominant reshaping mechanism.
Whether it does depends, in part, on the incompletely understood details of
surface processes operating at scales below the resolution of ground- or
space-based remote observations. One such process is the ``tangential YORP''
effect \citep{Gol12,Gol14}, arising from differential morning/afternoon heat
conduction across exposed surface features, roughly at the centimeter scale.
Owing to nighttime cooling, the temperature difference between the east and
west sides of such a feature is greater in the morning, leading to greater
conductive transport. The larger infrared flux radiated from the west side
then results in an eastward recoil force and an acceleration of the spin.
A second, related process is infrared ``beaming,'' caused by unresolved
surface roughness over a wide range of scales. Beaming directs the radiated
intensity slightly away from the mean surface normal, toward the direction
of the Sun. Again due to nighttime cooling, the effect is greater in the
morning \citep{Roz12}, enhancing the tangential recoil force in the westward
direction, and decelerating the spin. Which of these competing mechanisms
wins is a critical issue for future work.

\section{Conclusions}

The timescales over which YORP can alter the rotation period and the obliquity
of a small asteroid are not necessarily the same, because the corresponding
torques couple to different aspects of its shape. For an object close to axial
or reflection symmetry, the spin timescale can be an order of magnitude or more
longer than the obliquity timescale; hence YORP can reorient its spin axis to
low obliquity without substantially changing its spin rate. Reorientation
timescales for known top-shaped NEAs are $\sim 10^5$ to $10^6 \yr$, one to
two orders of magnitude shorter than their inner Solar System residence times.
Hence the observed low obliquities of these objects do not constitute evidence
that they were reshaped by YORP spin-up, because objects shaped by
any other process would also have been quickly reoriented to low obliquity.

Top shapes have not yet been established as a unique consequence of
YORP spin-up. Spin-driven generation of top-shaped
rubble piles has been shown to occur in a specific region of
parameter space, using a single numerical approach that neglects the 
unavoidable weakening of the YORP spin torque for near-symmetric shapes.
Determining whether this attractive explanation for top shapes is the
correct one will require significant effort, both to clarify the
relative importance of small-scale surface effects and to explore
the viability of competing mechanisms.

\bigskip
The author is grateful for helpful comments from colleagues
including Derek Richardson, Patrick Michel, Doug Hamilton, and Mangala
Sharma, and for support from NASA Planetary Geology \&
Geophysics grant NNX11AP15G. He also thanks the referees,
David Rubincam and David Vokrouhlicky, for constructive and helpful reviews.
This work has made use of NASA's Astrophysics Data System Bibliographic
Services.

\bibliography{asteroids}
\bibliographystyle{elsarticle-harv}

\end{document}